\documentclass[final,3p,times,twocolumn]{elsarticle}

\usepackage[english]{babel}

\usepackage[utf8]{inputenc}
\usepackage{physics}
\usepackage{listings}
\usepackage{color}
\usepackage{multirow} 
\usepackage{placeins}
\usepackage{graphicx}
\usepackage[T1]{fontenc}
\usepackage{soul}
\usepackage{amssymb}
\usepackage{amsmath}
\usepackage{hyperref}
\usepackage{float}
\usepackage{lipsum}

\graphicspath{{./Figs/}}
						
\journal{J. Magn. Magn. Mater.}

\begin{document}
\begin{frontmatter}
\title{Role of nanoparticle shape on the critical size for quasi-uniform ordering: from spheres to cubes through superballs.}

\author[USC]{Iago López-Vázquez
}
\ead{iago.lopez.vazquez@rai.usc.es}
\author[USC,iMATUS]{David Serantes
}
\ead{david.serantes@usc.gal}
\author[UB]{Òscar Iglesias
}
\ead{oscariglesias@ub.edu}
\affiliation[USC]{organization={Dpt. de Física Aplicada, Universidade de Santiago de Compostela},
            addressline={}, 
            city={Santiago de Compostela},
            postcode={15782}, 
            state={Galicia},
            country={Spain}}
\affiliation[iMATUS]{organization={Instituto de Materiais (iMATUS), Universidade de Santiago de Compostela},
            addressline={}, 
            city={Santiago de Compostela},
            postcode={15782}, 
            state={Galicia},
            country={Spain}}						

\affiliation[UB]{organization={Dpt. de Física de la Matèria Condensada, Universitat de Barcelona and IN2UB},
            addressline={c/ Mart\'{\i} i Franquès 1}, 
            city={Barcelona},
            postcode={08028}, 
            state={Catalunya},
            country={Spain}}

\begin{abstract}
The equilibrium states of single-domain magnetite nanoparticles (NPs) result from a subtle interplay between size, geometry, and magnetocrystalline anisotropy. 
In this work, we present a micromagnetic study of shape-controlled magnetite NPs using the superball geometry, which provides a continuous interpolation between spheres and cubes. 
By isolating the influence of shape, we analyze the transition from quasi-uniform (single-domain) to vortex-like states as particle size increases, revealing critical sizes that depend on the superball exponent $p$. 
Our simulations show that faceted geometries promote the stabilization of vortex states at larger sizes, with marked distortions in the vortex core structure.
The inclusion of cubic magnetocrystalline anisotropy, representative of magnetite, further lowers the critical size and introduces preferential alignment along the [111] easy axes. 
For isotropic shapes, the critical size for this transition increases with p, ranging from ~49 nm for spheres to ~56 nm for cubes, in agreement with experimental trends.
In contrast, the presence of slight particle elongation increases the critical size and induces another preferential alignment direction.
These results demonstrate that even small deviations from sphericity or aspect ratio significantly alter the magnetic ordering and stability of equilibrium magnetic states.







\end{abstract}




\end{frontmatter}

\section{Introduction}
Based on their promising uses for a variety of applications, ranging from magnetic recording \cite{chureemart2016model} to biomedicine \cite{soetaert2020cancer} or catalysis \cite{ovejero2021selective}, magnetic nanoparticles (MNPs) have been the subject of intense research attention during the last decades. Two key aspects drive this attention: firstly, the reduced size in itself, which would allow, for example, an increase in the areal density in heat-assisted magnetic recording \cite{john2017magnetisation}, or accurate control over cellular reactions by selectively triggering specific ion channels \cite{del2022magnetogenetics}. Secondly, the fact that, upon size reduction, peculiar magnetic properties arise which often differ from their bulk counterparts. Such originate by the change in relative importance of the energy terms at the reduced sizes: the dominance of exchange over magnetostatic energy at very small sizes results in coherent-like behaviour of the inner magnetic moments, so that the particles' magnetisation can be effectively described as that of a large \textit{supermoment}. Complementarily, the increasing fraction of moments at the particle surface, with different symmetry, results on enhanced role of the surface/shape effects \cite{skomski2003nanomagnetics,Iglesias_Book2021}.

Recent advances in NP synthesis have allowed the precise control of their geometry and dimensionality, enabling the production of faceted nanocubes, elongated nanorods, nanoplates, and other anisotropic architectures \cite{Pearce_NatRevChem2021, Nguyen_Nanom_2025} with high monodispersity \cite{Ma_ChemRev2023, Chang_ProgMater2024}. 
These advances have allowed fine-tuning of key physical parameters such as aspect ratio, surface faceting, and crystallinity \cite{Batlle_JMMM2022}, all of which are known to critically influence magnetic anisotropy and the stability of uniform magnetization states.

The particle shape constitutes a particularly key parameter to address due to its double key role, as it determines both the magnetic behavior of the individual particles and the collective dynamics of their ensembles. Thus, on the one hand, shape modulates the magnetostatic energy contributions and this leads to different single-particle properties \cite{roca2019design} and interparticle magnetization dynamics in NP assemblies \cite{gavilan2021size}. On the other hand, shape governs colloidal behaviour by affecting the geometrical packing, agglomeration \cite{donaldson2017nanoparticle}, surface electrostatics and interaction with the environment \cite{arciniegas2020unveiling}. We aim in this work to study, by means of a computational procedure, the role of particle shape on the threshold size for coherent/non-coherent behaviour of magnetite nanoparticles.


\begin{figure}[thp]
\centering
\includegraphics[width=0.9\columnwidth]{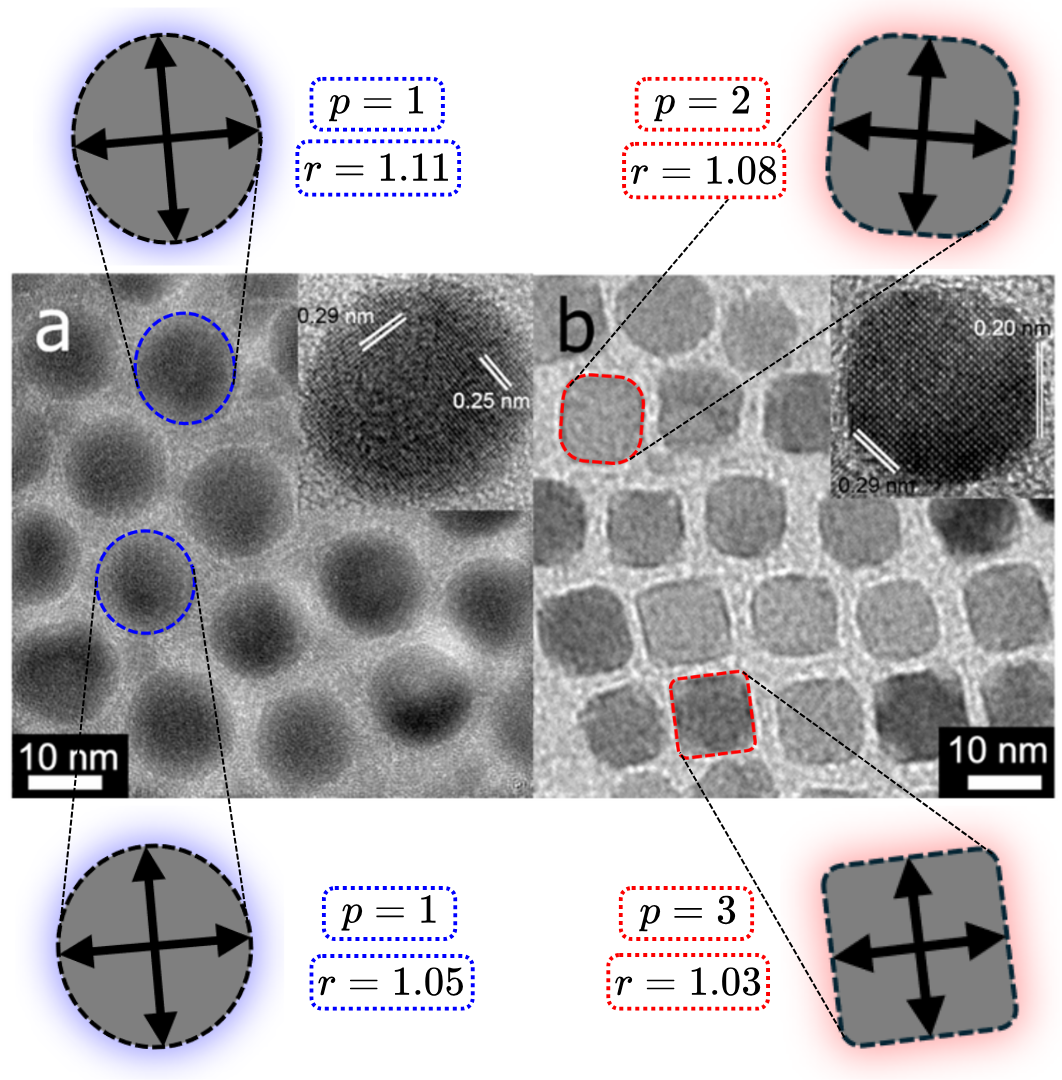}
\caption{\label{Fig_xperim_examples} Schematic representation of superellipsoidal shapes corresponding to different exponents $p$ and axial ratios $r = a/c$. It can be seen how small deviations from spherical or cubic symmetry give rise to more flattened or elongated shapes, similar to those observed experimentally. TEM images modified
from \cite{salazar2008cubic} with the permission of American Chemical Society.}
\end{figure}

A key motivation for the present study comes from our recent work \cite{Failde_Nanoscale2024}, in which we demonstrated that the hyperthermia efficiency of MNPs is highly sensitive not only to the magnitude but also to the symmetry of the magnetic anisotropy of the NPs. 
Building on these insights, the present study seeks a more comprehensive understanding of how subtle differences in shape can influence the equilibrium magnetic states, and thus the functional properties of real NP systems. Our starting point is the study of the role of the particle shape in the magnetic configuration of MNPs that are symmetric in the three spatial directions (i.e., same dimensions in the $x$, $y$, and $z$ axes), but different shapes. The motivation for selecting these shapes lies in the often inaccurate identification of “spherical particles” in experimental studies, where such particles typically do not exhibit a perfectly isotropic structure \cite{salazar2008cubic}. In this context, small variations in shape, combined with changes in the axial ratio, may offer a more realistic description of the shape polydispersity observed in real magnetic systems, as shown in Fig.\,\ref{Fig_xperim_examples}. 
Specifically, we will analyze the spontaneous magnetization of spheres, cubes, and intermediate superball shapes \cite{Disch_Nanoscale2013,Dresen2021neither}. 
Spontaneous magnetization is selected as a characteristic parameter because it reflects the appearance of different magnetic states as the particle size increases \cite{gavilan2021size, lopez2021mapping}. 

The work is organised as follows: in Secs.\,\ref{sec_pysical_model} and \ref{Sec_Computational} the physical model and the computational details are described, respectively. Then, the following sections correspond to results and discussion of different physical conditions. In Sec.\,\ref{sec_no-Kc} we first investigate the role of exchange \textit{vs.} magnetostatic energy in the ideal scenario  where magnetocrystalline anisotropy is absent, which is introduced in Sec.\,\ref{sec_with-Kc}. Then, in Section \ref{sec_aspect-ratio} the role of the aspect ratio is investigated. Finally, in Sec.\, \ref{Sec_Vortex} different vortex configurations are analysed. The results of the work are summarised in Sec.\, \ref{sec_conclusions}.

\section{Physical model}\label{sec_pysical_model}
\begin{figure}[thp]
\centering
\includegraphics[width=0.9\columnwidth]{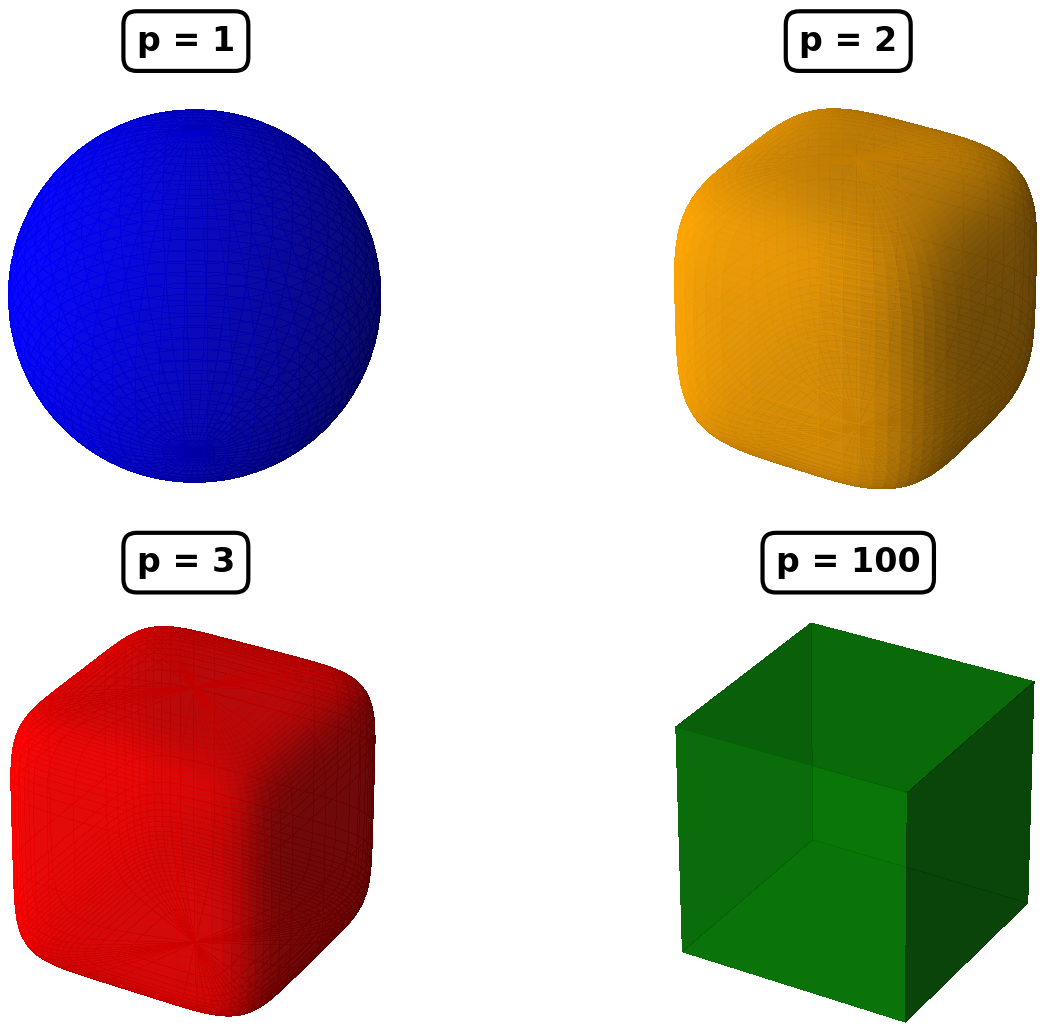}
\caption{\label{Fig_scheme-shapes} Schemes showing the different geometries studied in this work: $p= 1$ (sphere), $p= 2$, $p= 3$ and $p= 100$ (cube).}
\end{figure}

Based on the interest for biomedical applications, we will focus our study on magnetite, considering the values of the characteristic parameters in bulk. To consider different particle shapes, we have used the equation of the superellipsoid \cite{donaldson2017nanoparticle,Dresen2021neither}:
\begin{equation}\label{Eq_Superball}
\left(\frac{x}{a}\right)^{2p}+\left(\frac{y}{b}\right)^{2p}+\left(\frac{z}{c}\right)^{2p}\leq 1\ .
\end{equation}
Considering the same dimensions along the three axes, $a=b=c$, increasing the value of the $p$ parameter in the exponent from $1$, allows one to continuously tune the NP shape from a perfect sphere to a cube, passing through intermediate cases that resemble a cube with round corners and edges, more similar to the shapes typically observed synthesized magnetite NPs. 
Examples of these shapes are shown in Fig.\,\ref{Fig_scheme-shapes}, of 4 different particles with the same volume.

It is important to emphasize that, when comparing results for different NP shapes varying $p$, we will consider the same magnetic volume, $V$, so that there is no difference on the associated magnetization and colloid characteristic timescales, which are directly proportional to $V$ \cite{balakrishnan2020exploiting}
(see, for example, \cite{coffey2020thermal} for a detailed description of the characteristic timescales for magnetization reversal, rotation, and diffusion). 
In this way, we can focus on the effect of shape only on the magnetic behavior.
 
Nevertheless, for the sake of simplicity, we will in general refer to particle size of any shape in terms of the equivalent cube side $L$. Keeping this equivalence in mind, in our study we will consider particle sizes from $L=40$ nm and above, as critical sizes for magnetite particles have been reported to be around $49$ nm for spheres \cite{petracic2010superparamagnetic}, and around $54$ and $56$ nm for octahedral \cite{vereda2008synthesis} and cube \cite{usov2018magnetic} shapes; our own simulations suggest that even the presence of an oxidized layer will not modify the critical size \cite{lopez2021mapping}. However, there are works indicating that the critical size may be significantly larger \cite{li2017correlation}. We hope our computational study, with well-controlled parameters, may contribute to shed some light on this subject of the critical size of magnetite nanoparticles.

\section{Computational details}\label{Sec_Computational}

To study the role of the particle shape on the magnetic properties, we have performed a micromagnetic study based on the Object Oriented MicroMagnetic Framework software (OOMMF) \cite{oommf} to find the zero temperature equilibrium configurations of the NPs, using the energy minimization driver \textit{Oxs MinDriver} with a stopping criterion \textit{stopping m$\times$H$\times$m}=0.1. 

To ensure that this convergence criterion was sufficient and that the final states did not depend on the initialization, most cases were also simulated starting from different initial configurations, including random orientations, uniform alignment along a specific direction and pre-imposed vortex-like states, consistently converging to the same equilibrium configurations.
%

The parameters characterizing the different energy terms are those corresponding to magnetite: exchange stiffness constant $A= 1.1\times 10^{-11}$ J/m, saturation magnetization $M_s= 4.8\times 10^5$ A/m, and cubic anisotropy constant $K_c=-1.1\times 10^{4}$ J/m$^3$.   
In all cases, we have used cubic discretization cells of side $1$ nm as the basic building block, which
is well below the exchange length of magnetite $L_{ex}\simeq 5.4$ nm, and small enough to resolve the kind of magnetic order to be studied for the NP size range that we will consider. We additionally verified the stability of the results by performing tests with smaller cell sizes, 
which produced the same equilibrium states. For this reason, and to keep the computational 
cost manageable for larger MNPs, we present only the results obtained with the 1 nm 
discretization throughout the manuscript.

We will focus on $4$ particular geometric shapes as pictured in Fig.\,\ref{Fig_scheme-shapes} corresponding to $p=1, 2, 3, 100$. 
Since we want to compare magnetic configurations of NPs of similar sizes, the lateral size of the superballs will be varied so that their volume corresponds to that of a cubic NP with edge $L$. 
In order to keep the volume of the NPs constant while changing $p$, we have calculated the equivalent cube sizes of superballs of index $p$ by using the following analytical expression \cite{jiao2009optimal} for the volume of a superball of radius $a$
\begin{equation}
    V_{sb}(p,a)= \frac{2a^3}{p^2} \frac{\Gamma\left(1+\frac{1}{2p}\right) \Gamma^2\left(\frac{1}{2p}\right)}{\Gamma\left(1+\frac{3}{2p}\right)}\ ,
\label{Eq_VolSuperball}
\end{equation}
where 
$\Gamma(x)$ is the Gamma function.

Equating this expression to that of a cube of side $L$, we obtain the equivalent superball radii, some of which are given in Table \ref{Tab1}.

\begin{table}[h]
\centering
\begin{tabular}{|c|c|c|c|}
\hline
$L_{cube}(nm)$ & $L_{p=3}(nm)$ & $L_{p=2}(nm)$ & $L_{sphere}(nm)$ \\ \hline
40             & 41.40                    & 42.91                    & 49.63           \\ \hline
50             & 51.75                    & 53.63                    & 62.04           \\ \hline
60             & 62.10                    & 64.36                    & 74.44           \\ \hline
80             & 82.80                    & 85.81                    & 99.25           \\ \hline
\end{tabular}
\caption{Values of the NP size having the same volume as a cube ($p = 100$) of side $L_{cube}$ for a superball with $p=3$, $p=2$ and a sphere ($p=1$).}
\label{Tab1}
\end{table}

\section{Equilibrium magnetization configurations of NPs with no anisotropy}\label{sec_no-Kc}
We will start by considering the ideal case in which only exchange and magnetostatic energy contributions are taken into account. In this case, the only source of anisotropy comes from the demagnetizing field contribution due to the NP shape. The results should be representative of NPs of soft materials such as permalloy or nickel.

\begin{figure}[thp]
\centering
\includegraphics[width=0.95\columnwidth]{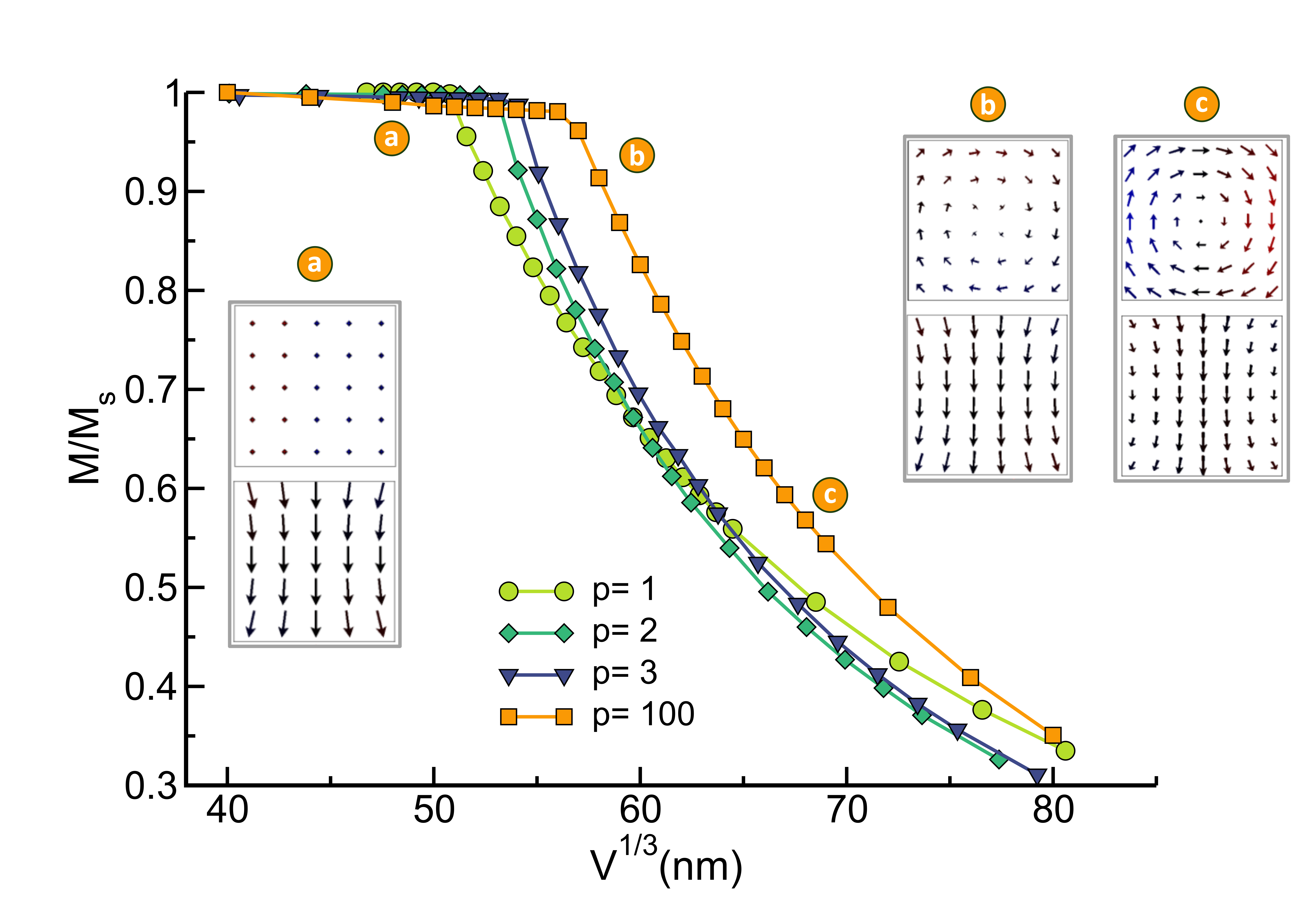}
\caption{\label{Fig_7} Size dependence of the normalized equilibrium magnetization modulus of NPs with no anisotropy and different shape exponents $p$ corresponding to spherical ($p= 1$, circles), superballs ($p= 2, 3$, rhombus and triangles) and cubic ($p= 100$, squares) shapes. 
The insets show schematic snapshots of central planes perpendicular to the $x$ and $y$ axes (upper and lower subpanels, respectively) of typical magnetic configurations of cubic NPs: (a) flower state for $L= 48$ nm; vortex state for (b) $L= 58$ nm and (c) $L= 68$ nm.}
\end{figure}

The magnetization modulus for the equilibrium configurations of a NP with different shape exponents ($p=1, 2, 3, 100$) is shown in Fig.\,\ref{Fig_7} as a function of the side length $L=V^{1/3}$ for sizes between $40$ and $80$ nm. 

In Fig.\,\ref{Fig_7} it is observed that for small sizes, the magnetization is close to saturation (ferromagnetic (FM) state) as expected, with all the magnetic moments aligned parallel to one of the symmetry axes. 
Increasing the size, small deviations from collinearity induced by the demagnetizing field appear, and the magnetic configurations correspond to the so-called flower state \cite{Aharoni_JAP1980}, in which the magnetic moments at the topmost layers and sides of the NP progressively deviate from the symmetry axis, as can be seen in panel (a) of Fig.\,\ref{Fig_7}.  
For NP volumes above a critical value ($V_c$), there is a transition to a vortex state with a FM alignment of the moments pointing near the symmetry axis of the NP (vortex core) and those closer to the surface circulating around the symmetry axis. This rotational configuration decreases the magnetostatic energy at the expense of an increase in exchange energy.
The vortex core region shrinks when the NP size is increased and the moments in the rotational region progressively rotate toward the plane transverse to the core axis, causing the magnetization decrease observed in Fig.\,\ref{Fig_7}.
Intermediate characteristic states are not formed, as indicated by the smoothness of the curves near $L_c$.

The transition between the flower and vortex states is progressive: as the particle size is increased, the magnetic moments closer to the sides of the flower state continuously reorient towards the $yz$ plane perpendicular to the symmetry axis, while those at the center of the particle remain pointing along this axis.
We have verified by direct inspection of the configurations that the vortex core is always perpendicular to one of the Cartesian axes. We defer a more in depth characterization of the magnetic moment configurations to Sec.\ref{Sec_Vortex}, where results including crystalline anisotropy will be analysed.

The critical size $L_c$ for the stability of quasi-uniform configurations has been systematically determined by performing two independent linear fits: one to the data points in the quasi-uniform regime, and another to the first five points in the vortex regime where $M/M_s < 0.95$. The intersection of these two fitted lines was then taken as the critical size for each case. The uncertainties obtained with this methodology are all below 0.1 nm, which is smaller than the symbol size used in the figures and therefore not visible.

As can be seen from the data presented in Fig.\,\ref{Fig_7}, $L_c$ is smaller for spherical NP ($p=1$) than for cubic ones ($p=100$) and increases gradually as the superball exponent increases. In our simulations, we obtain approximate values of $V_c^{1/3}=51$ nm for the sphere ($p=1$) and $56$ nm for the cube ($p=100$), which are in very good agreement with previously reported values of $49$ nm for spheres \cite{petracic2010superparamagnetic} and $56$ nm for cubes \cite{usov2018magnetic}.  
Expressed in units of the exchange length of magnetite ($L_{ex}\simeq 8.7$ nm), we obtain $L_c \simeq 5.6\,L_{ex}$ for the sphere and $L_c \simeq 6.4\,L_{ex}$ for the cube. The first value is somewhat smaller than the theoretical prediction of $7.2\,L_{ex}$ for a perfect sphere reported in 
Ref.~\cite{di2012generalization}. This discrepancy may be mainly attributed to two factors: (i) the different discretization schemes, namely a continuous model in Ref.~\cite{di2012generalization} versus our finite-difference approach (especially concerning surface effects), and (ii) the fact that in our case the transition is studied from a quasi-uniform state rather than from a fully aligned configuration, whereas in Ref.~\cite{di2012generalization} the magnetization is assumed to be spatially constant.

For NP sizes near, but greater than $L_c$, the magnetization of the equilibrium configurations is higher for NP with greater $p$. However, for sizes $L> 68$ nm the equilibrium magnetization of spherical NP become larger than that of superballs with rounded corners and seems to converge to the one of cubic NP for sizes near $80$ nm.

\begin{figure}[H]
\centering
\includegraphics[width=1.0\columnwidth]{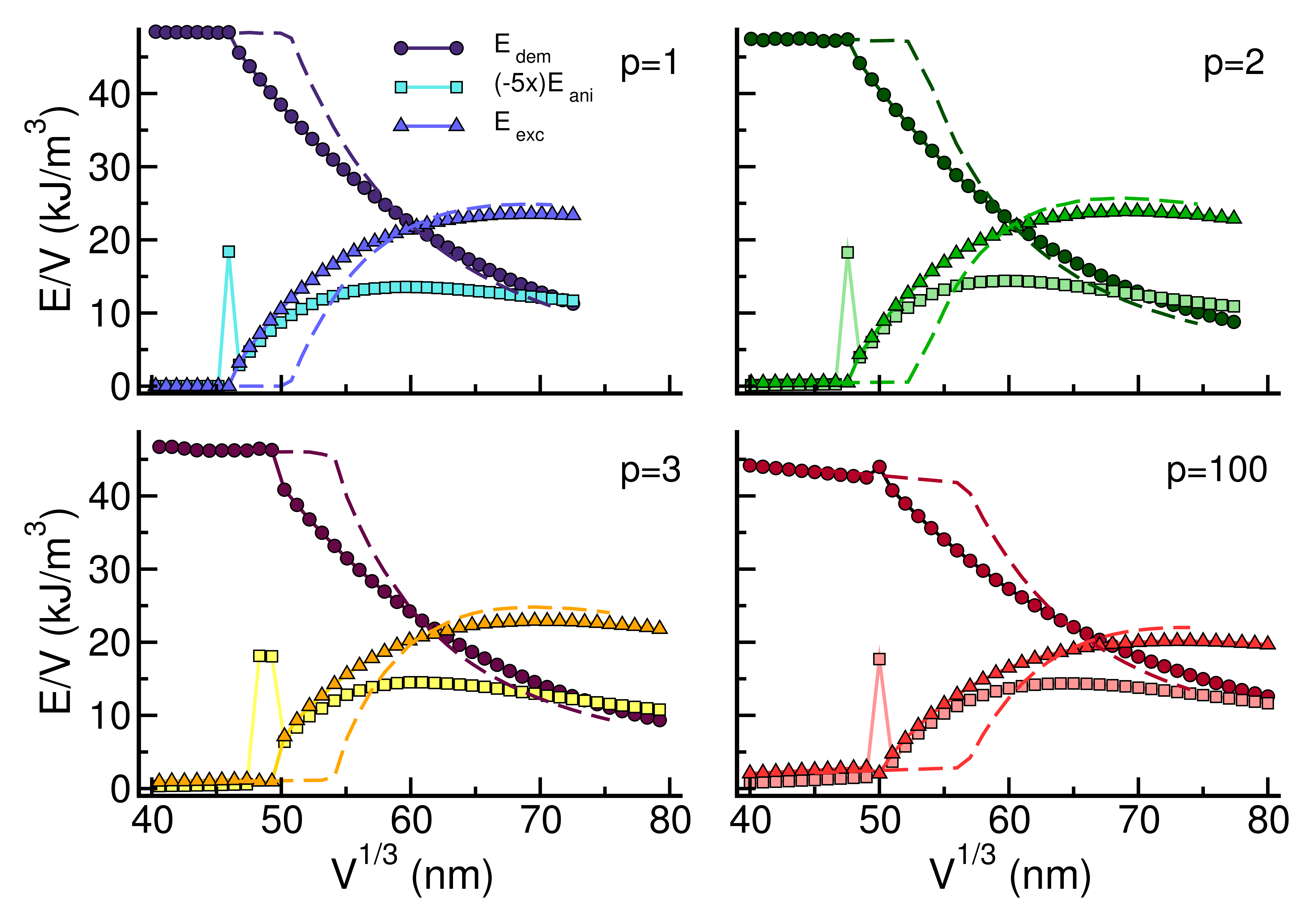}
\caption{\label{Fig_Energies} Energy contributions of the equilibrium configurations of NPs without (dashed lines) and with cubic magnetocrystalline anisotropy (symbols) as a function of the equivalent NP size $V^{1/3}$. $E_{\text dem}$ (circles), $E_{\text exc}$ (triangles), and $E_{\text ani}$ (squares) stand for the demagnetizing, exchange and anisotropy energies, respectively. $E_{\text ani}$ has been rescaled by a factor $-5$ for better visibility. Results are shown for $4$ different shapes corresponding to $p= 1, 2, 3, 100$ as indicated in each subpanel.}
\end{figure}

To better understand the size dependence of the magnetization, it is instructive to inspect the changes in the different energy contributions to the total energy as a function of the NP size, as depicted in Fig. \ref{Fig_Energies} by dashed lines for $4$ values of the shape exponent $p$.
In the flower state, the demagnetizing energy $E_{\text{dem}}$ decreases and the exchange energy increases with increasing shape parameter. Both energies show a linear dependence on the NP size, independently of $p$. Moreover, the deviation from the FM state is more pronounced for cubes than for spheres. 
For the smallest simulated sizes, $E_{\text dem}$ converges to that of a state with uniform magnetization \cite{Aharoni_JAP1980,Coey20101}: 
$E_{\text dem}= \mu_0 M_s^2/6 = 4.82\times 10^{4}$ J/m$^3$ for a sphere and a cube ($p= 1, 100$).
In the vortex state, $E_{\text{dem}}$ has a fast monotonic decrease with increasing NP size that seems to tend to a constant value. 

The exchange energy $E_{\text{exc}}$ first has a fast and non-linear increase and reaches a maximum for sizes around $70$ nm (although for cubic NP the maximum is not reached within the considered size range), and then slowly decreases.
States with equal values of $E_{\text{dem}}$ and $E_{\text{exc}}$ occur for a NP size that depends on $p$ (from $60$ nm for $p=1$, up to $66$ nm for $p= 100$). However, they all are characterized by a vortex configuration and do not seem to correspond to a particular kind of magnetic order different from that of NPs with similar sizes.

\section{Equilibrium magnetization configurations of NPs with cubic anisotropy}\label{sec_with-Kc}
\begin{figure}[!ptb]
\centering
\includegraphics[width=1.0\columnwidth]{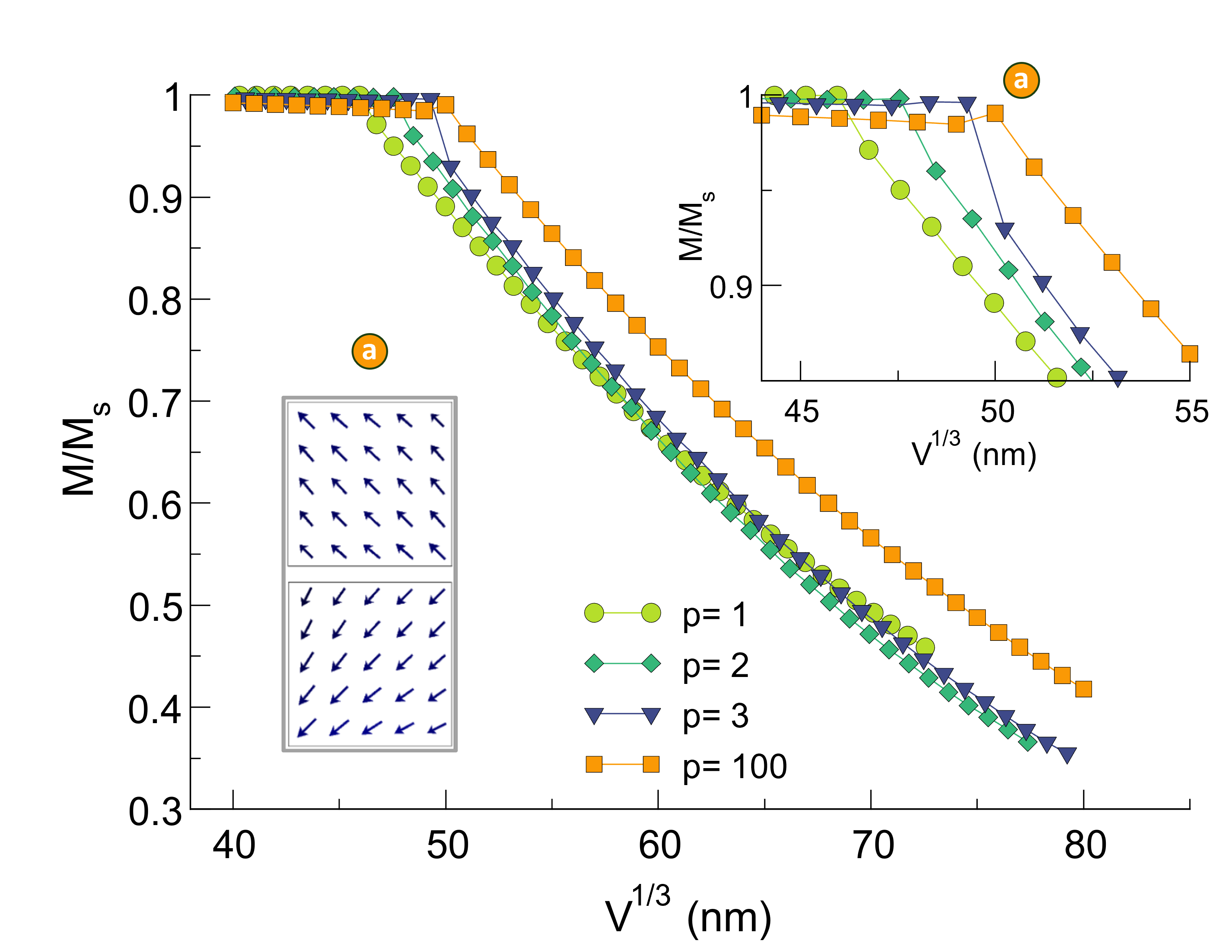}
\caption{\label{Fig_15} Size dependence of the normalized equilibrium magnetization of magnetite NPs with cubic anisotropy and different shape exponents ($p= 1, 2, 3, 100$) corresponding to shapes from spherical ($p= 1$), superballs ($p= 2, 3$) to cubic ($p= 100$). 
The inset shows an expanded region of the main panel around the critical size.
The subpanels in the left display schematic snapshots of central planes perpendicular to the $x$ and $y$ axes (upper and lower subpanels, respectively) of the magnetic configuration of a cubic NP with the critical size, marked with an orange circle.
}
\end{figure}

To investigate the influence of magnetocrystalline anisotropy on the equilibrium magnetization configurations, we incorporate anisotropy energy into our model. As discussed Sec.\,\ref{Sec_Computational}, we focus on the case of magnetite, which exhibits cubic magnetocrystalline anisotropy with a negative anisotropy constant. The critical sizes for the flower-to-vortex transition were obtained following the same methodology outlined in the previous subsection, and the corresponding results are presented in Fig.\,\ref{Fig_15} for $p= 1, 2, 3, 100$.
A direct comparison with the zero-anisotropy case (see Fig.\,\ref{Fig_7}) clearly shows that inclusion of anisotropy results in a reduction in the critical volumes across all shapes. 
This shift can be understood by examining the energy balance, as will be shown the following.
In addition, the evolution of the reduced magnetization with increasing NP size exhibits a notable change in behavior.
In the presence of anisotropy, $M/M_s$ decreases more gradually compared to the isotropic case, displaying an almost linear trend throughout the vortex regime.

A closer look at the behavior of $M/M_s$ near the critical size for the flower-to-vortex transition in a cubic NP ($p= 100$) reveals a subtle upturn of the magnetization near $L_c= 50$ nm (see the inset in Fig.\,\ref{Fig_15}). Inspection of the corresponding equilibrium magnetization configuration at this point (labelled $a$ in Fig.\,\ref{Fig_15}) indicates that, at this size, the magnetic moments preferentially align along one of the four easy axis directions, which, in the case of negative cubic anisotropy as in magnetite, corresponds to a cube diagonal, as schematized in the inset of Fig.\,\ref{Fig_15}. 
This intermediate state between flower and vortex configurations, characterized by a uniform alignment along a cube diagonal, also arises in superballs with rounded vertices and in spherical NPs, although it is not explicitly shown in Fig.\,\ref{Fig_15}.
Only for $p=3 $, the transition from flower to vortex is more abrupt (as seen from the sharp drop in magnetization marked by triangle symbols in Fig.\,\ref{Fig_15}). As in the zero anisotropy case, no diagonal vortex states are stabilized in the studied size range for any shape \cite{Bonilla_JMMM2017,GuoNano_Lett2022}.

This transition can be further understood by analyzing the contributions of the individual energy terms to the total energy.
The demagnetizing $E_{\text{dem}}$, exchange $E_{\text{exc}}$ and anisotropy $E_{\text{ani}}$ contributions to the total energy (with $E_{\text{ani}}$ scaled by a factor $-5$ for better visibility) are plotted as a function of the NP size in Fig.\,\ref{Fig_Energies}, for $4$ values of the shape exponent $p$. 
Both $E_{\text{dem}}$ and $E_{\text{exc}}$ trends analogous to those observed in the absence of anisotropy, but with smoother size dependence in the vortex regime, attributable to the contribution of the negative anisotropy term. As expected, $E_{\text{ani}}$ vanishes for the FM and flower states.
However, despite being approximately $5$ times smaller than $E_{\text{exc}}$, $E_{\text{ani}}$ increases sharply at the flower-to-vortex transition for all $p$, clearly indicating magnetic configurations that are predominantly aligned along one of the $[111]$ easy axes of magnetite, which minimize magnetocrystalline energy at the expense of only a marginal increase in $E_{\text{dem}}$.

To summarize this discussion about the role of particle shape, in Table~\ref{tab:critical_sizes} we  represent the approximated critical sizes obtained in this work for different geometries, together with previously reported values in the literature.

\begin{table}[h]

\centering
\begin{tabular}{|c|c|c|c|}
\hline
\multirow{3}{*}{\textbf{Shape ($p$)}} & 
\multicolumn{2}{c|}{\textbf{This work}} & 
\multirow{2}{*}{\textbf{Reported}} \\ \cline{2-3}
 & $K_c = 0$ & $K_c \neq 0$ &  \\ \cline{2-4}
 & $V_c^{1/3}$ (nm) & $V_c^{1/3}$ (nm) & $V_c^{1/3}$ (nm) \\ \hline
$p=1$   & 51 & 46 & 49 \cite{petracic2010superparamagnetic} \\ \hline
$p=2$   & 54 & 48 & \multirow{2}{*}{54 \cite{vereda2008synthesis}} \\ \cline{1-3}
$p=3$   & 53 & 49 &  \\ \hline
$p=100$ & 56 & 50 & 56 \cite{usov2018magnetic} \\ \hline
\end{tabular}
\caption{Critical sizes $V_c^{1/3}$ for magnetite nanoparticles of different geometries. Results are shown for $K_c=0$ and $K_c\neq 0$ and compared with previously reported values.}
\label{tab:critical_sizes}
\end{table}

\section{Influence of aspect ratio on critical sizes}\label{sec_aspect-ratio}
Since experimentally synthesized NPs typically exhibit some deviation from the ideal aspect ratio 
$r=1.0$, in this section, we analyse the effects of varying this parameter by elongating the NPs along one of the symmetry axes of the superball shape.

\begin{figure}[H]
\centering
\includegraphics[width=0.9\columnwidth]{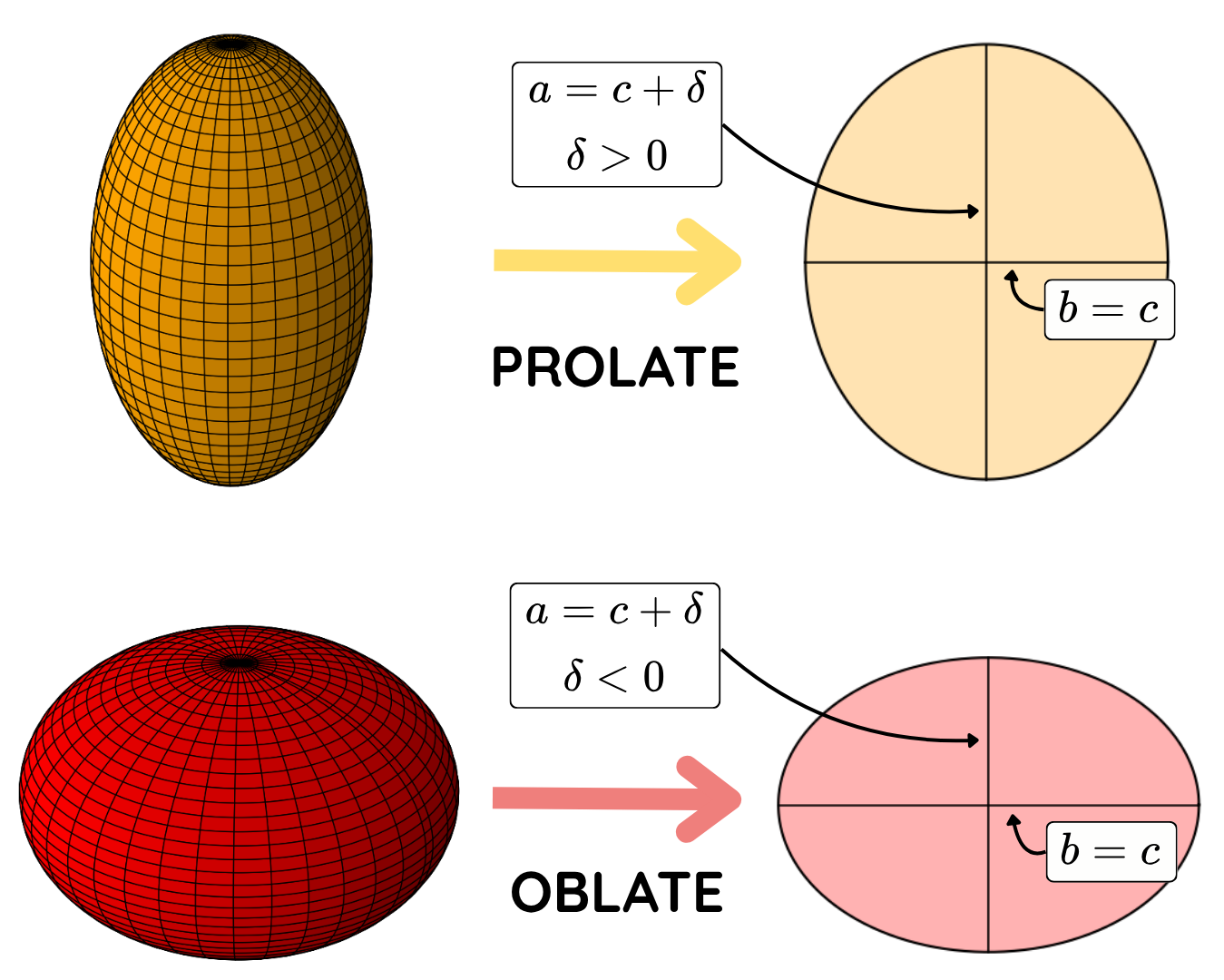}
\caption{Schematic representation of elongated (prolate, $\delta>0$) and compressed 
(oblate, $\delta<0$) superellipsoids used in the aspect ratio analysis. 
Here $a=c+\delta$ and $b=c$.}
\label{Fig_prolate_oblate}
\end{figure}

For this purpose, a generalization of Eq. \ref{Eq_Superball} is necessary. This can be achieved by using the definition of a superellipsoid surface as the intersection of Lam\'e curves as presented in \cite{Jaklic2000}
\begin{equation}\label{Eq_Superball2}
\left( \frac{x}{a} \right)^{\frac{2}{\epsilon_1}} + 
\left( \left( \frac{y}{b} \right)^{\frac{2}{\epsilon_2}} + 
\left( \frac{z}{c} \right)^{\frac{2}{\epsilon_2}} \right)^{\frac{\epsilon_2}{\epsilon_1}}
\leq 1\ .
\end{equation}
Here, $\varepsilon_1, \varepsilon_2$ are related to the exponents of the original superellipses. Eq. \eqref{Eq_Superball} is recovered when $\varepsilon_1,=\varepsilon_2=1/p$. 
Considering shapes with $b=c$, we will vary the semiaxis along the $x$ direction according to $a=c+\delta$ where $\delta$ corresponds to the elongation ($\delta>0$) or contraction ($\delta<0$) of the original shape in nm, generating superellipsoids with increasing prolate or oblate shapes (Fig.~\ref{Fig_prolate_oblate}).

In this way, we will generate shapes with different aspect ratios $r= a/c$ and volumes given by the formula \cite{Jaklic2000} 
\begin{multline}
   \frac{V_{sb}}{abc}
		=
		2\varepsilon_1\varepsilon_2 
		B\left(\frac{\varepsilon_1}{2},\varepsilon_1+1\right)
		B\left(\frac{\varepsilon_2}{2},\frac{\varepsilon_2+2}{2}\right)\\
		= \frac{2}{p^2}
		B\left(\frac{1}{2p},\frac{2p+1}{2p}\right)
		B\left(\frac{1}{2p},\frac{p+1}{p}\right)
\ ,
\label{Eq_VolSuperball_2}
\end{multline}
where $B(x,y)=\Gamma(x)\Gamma(y)/\Gamma(x+y)$ is the beta function.
As mentioned in Sec.\,\ref{Sec_Computational}, we will continue to consider NPs with the same magnetic volume when changing $p$ for a given value of $r$ As mentioned in Sec.~\ref{Sec_Computational}, throughout this work the particle size is always expressed in terms of $V^{1/3}$. When elongation is introduced ($r \neq 1$), the semiaxis $a$ is modified and the corresponding volume is calculated from Eq.~\eqref{Eq_VolSuperball_2}. In this way, the results for each geometry and elongation are consistently represented as $M/M_s$ vs.~$V^{1/3}$. Strictly speaking, each $V^{1/3}$ value corresponds to a slightly different axial ratio $r$, but in the analysis we refer to the effective $r$ associated with the critical volume for each elongation. This procedure ensures that comparisons between shapes and elongations are made on the basis of the effective magnetic volume.

\begin{figure}[H]
\centering 
\includegraphics[width=1.0\columnwidth]{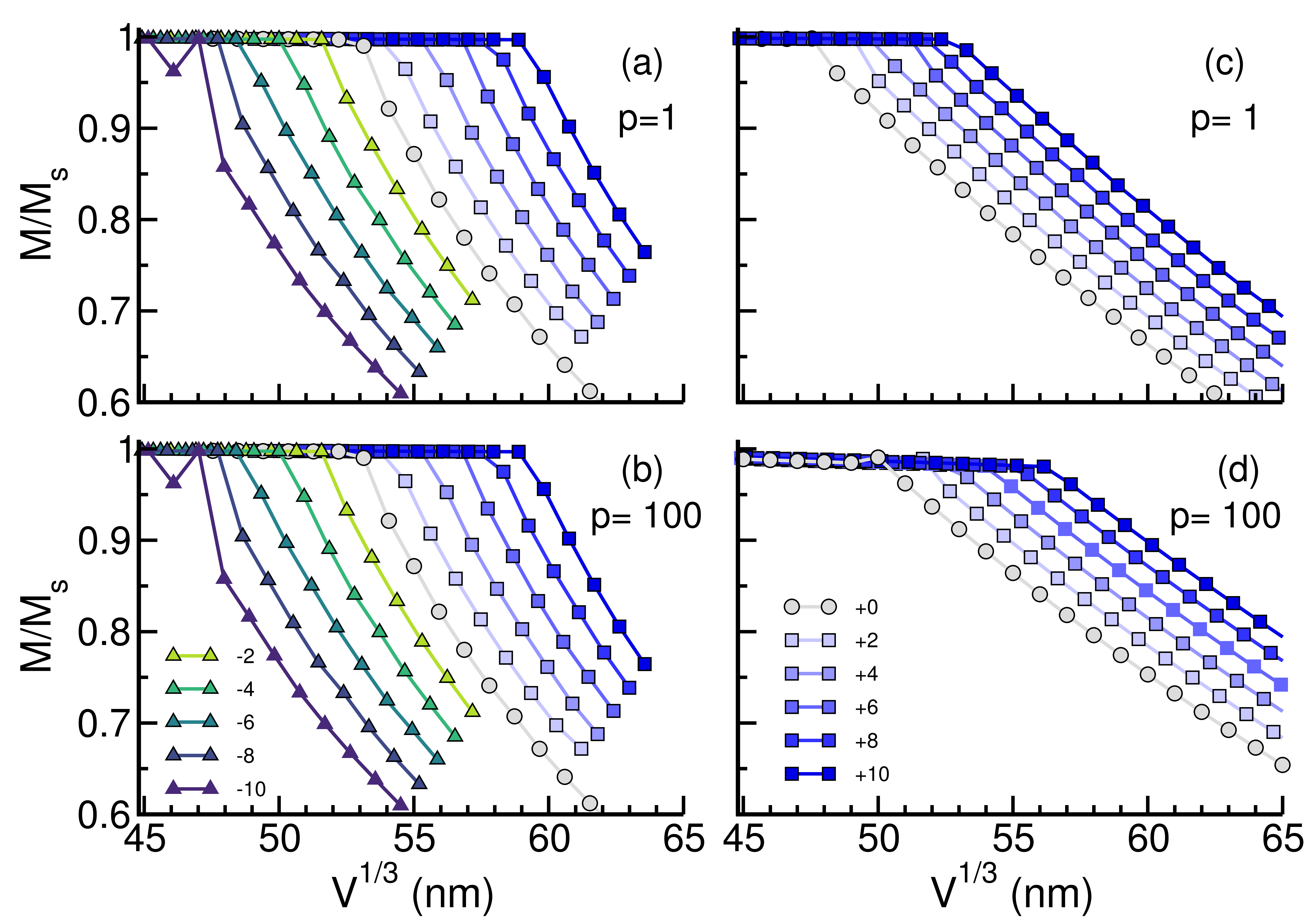}
\caption{\label{Fig_10} Size dependence of the normalized equilibrium magnetization for magnetite NPs without (left panels (a) and (b)) and with magnetocrystalline anisotropy (right panels (c) and (d)). The different curves correspond to superball shapes with different lengths of the long symmetry axis $a=b\pm \delta$, where $\delta$ is the elongation or contraction along one of the symmetry axes in nm as indicated in the legends. The gray circles correspond to the non-elongated NPs. 
}
\end{figure}
The results for the equilibrium magnetization configurations are presented in Fig.\,\ref{Fig_10} for two representative values of the shape exponent $p=1$ and $p= 100$.
The data correspond to the cases without anisotropy (panels (a,b)) and with anisotropy included (panels (c,d)), across a range of elongation parameters $\delta=2,\dots, 10$ nm. In the former case, oblate shapes ($\delta<0$) have also been considered.

In all scenarios, increasing the aspect ratio leads to an increase in the critical size, regardless of the shape exponent $p$.
This trend stems from the uniaxial shape anisotropy induced by elongation, which dominates over the exchange energy contribution on a wide range of volumes and favours quasi-uniform states with the magnetization aligned along the major axis $a$. 
In contrast, for increasingly oblate shapes, the critical size decreases. 
In this case, uniaxial shape anisotropy promotes magnetization lying in the $yz$ plane, facilitating the formation of vortex states revolving around the $x$ direction at smaller volumes.

Consequently, just above $V_c$, for a given volume, isotropic NPs have a higher magnetization component along the elongated axis compared to anisotropic ones.
However, this trend reverses for NPs at larger sizes. 
Because $V_c$ is smaller and the magnetization decays more rapidly in the isotropic case, the magnetization vs. $V_c^{1/3}$ curves for the two cases intersect at some point. 
Beyond this crossover size, NPs with anisotropy exhibit a higher magnetization than that of their isotropic counterparts, regardless of the aspect ratio.

Once the vortex regime is reached, both cases exhibit a linear decrease in magnetization with increasing size, and this trend holds for all values of $p$ considered, although the decrease is more pronounced in the absence of anisotropy (compare panels (a) and (c) in Fig.\,\ref{Fig_10}).
\begin{figure}[H]
\centering
\includegraphics[width=0.9\columnwidth]{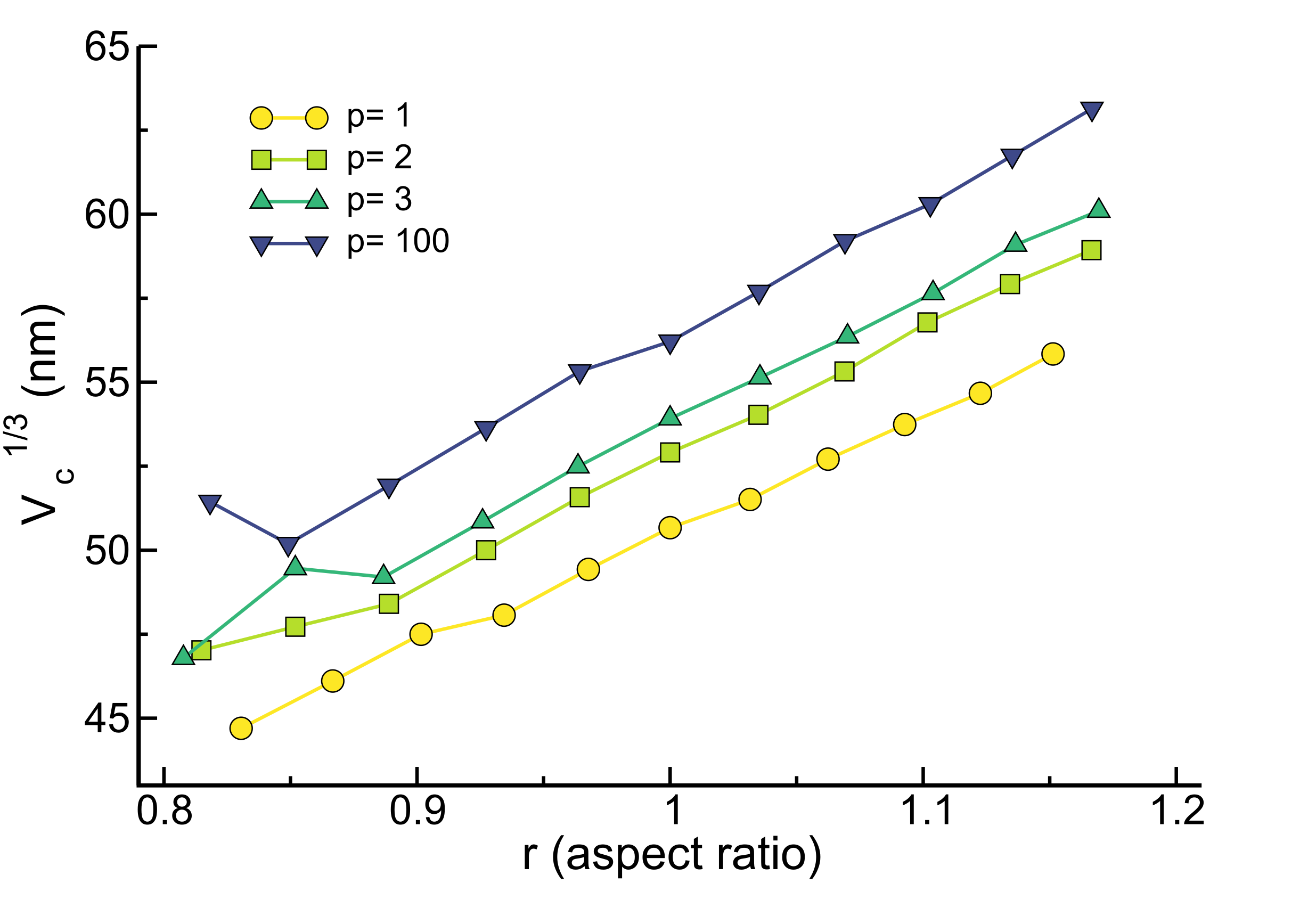}
\caption{Critical size of NPs without cubic crystalline anisotropy as a function of the aspect ratio $r$. Results are shown for four different shapes corresponding to $p= 1, 2, 3, 100$ as indicated in the legend. \label{Fig_13}}
\end{figure}

In Fig.\,\ref{Fig_13}, which shows the dependence of the critical size on the aspect ratio for NPs without crystalline anisotropy and different shapes, the influence of the NP shape can be clearly appreciated: as the NP shape transitions from a sphere ($p=1$) to a cube ($p=100$), the critical size increases.
This can be attributed to the higher demagnetizing energy density near $V_c$ in the cubic case, and this effect is maintained regardless of whether anisotropy is included in the model.

\begin{figure}[H]
\centering
\includegraphics[width=1.0\columnwidth]{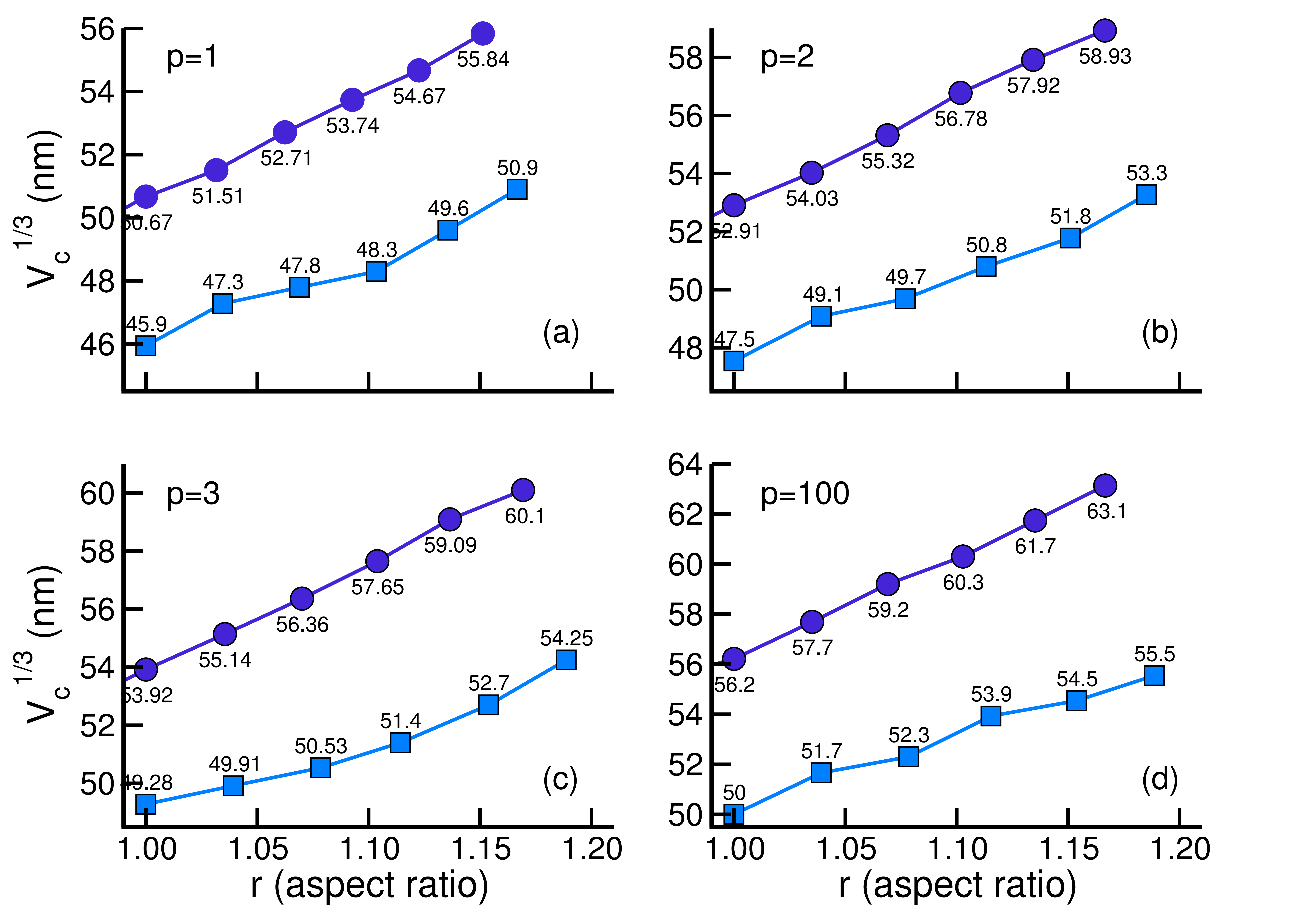}
\caption{Critical size of magnetite NPs as a function of the aspect ratio $r$. Results are shown for four different shape corresponding to $p= 1, 2, 3, 100$ as indicated in the legends. For each $p$, the results (not) taking into account anisotropy are plotted using (violet circles) blue squares.  \label{Fig_Resumen}}
\end{figure}

The trends described above are common to both the isotropic and anisotropic cases, as illustrated more clearly in Fig.\,\ref{Fig_Resumen}, where the cases with and without anisotropy are represented for each NP shape studied.
The inclusion of cubic anisotropy leads to a lower $V_c$ compared to the isotropic case. 
This reduction arises because cubic anisotropy counteracts the uniaxial shape-induced anisotropy of the system, favoring reorientation of the magnetic moments toward the easy axes along the cube diagonals to minimize the total energy. As a consequence, the magnetic moments lose their coherent alignment at smaller volumes.

\section{Analysis of configurations with vortex states}\label{sec_analysis-vortices}
\label{Sec_Vortex}

In this final section, we present further insights into the equilibrium configurations obtained above the critical size, with a focus on the detailed structure of the resulting vortices.
Specifically, we investigate how the vortex morphology evolves with increasing NP size and examine the influence of both the superball shape and the elongation of the NPs on the vortex characteristics. 
Due to the large number of discretization cells required in this size range, we begin by presenting representative snapshots of magnetic configurations across different planes perpendicular to the symmetry axes. 
These visualizations display cone or arrow glyphs indicating the local magnetization direction, with color scales representing the intensity of one of the magnetization components.

\begin{figure*}[!ptb]
\centering
\includegraphics[width=0.9\textwidth]{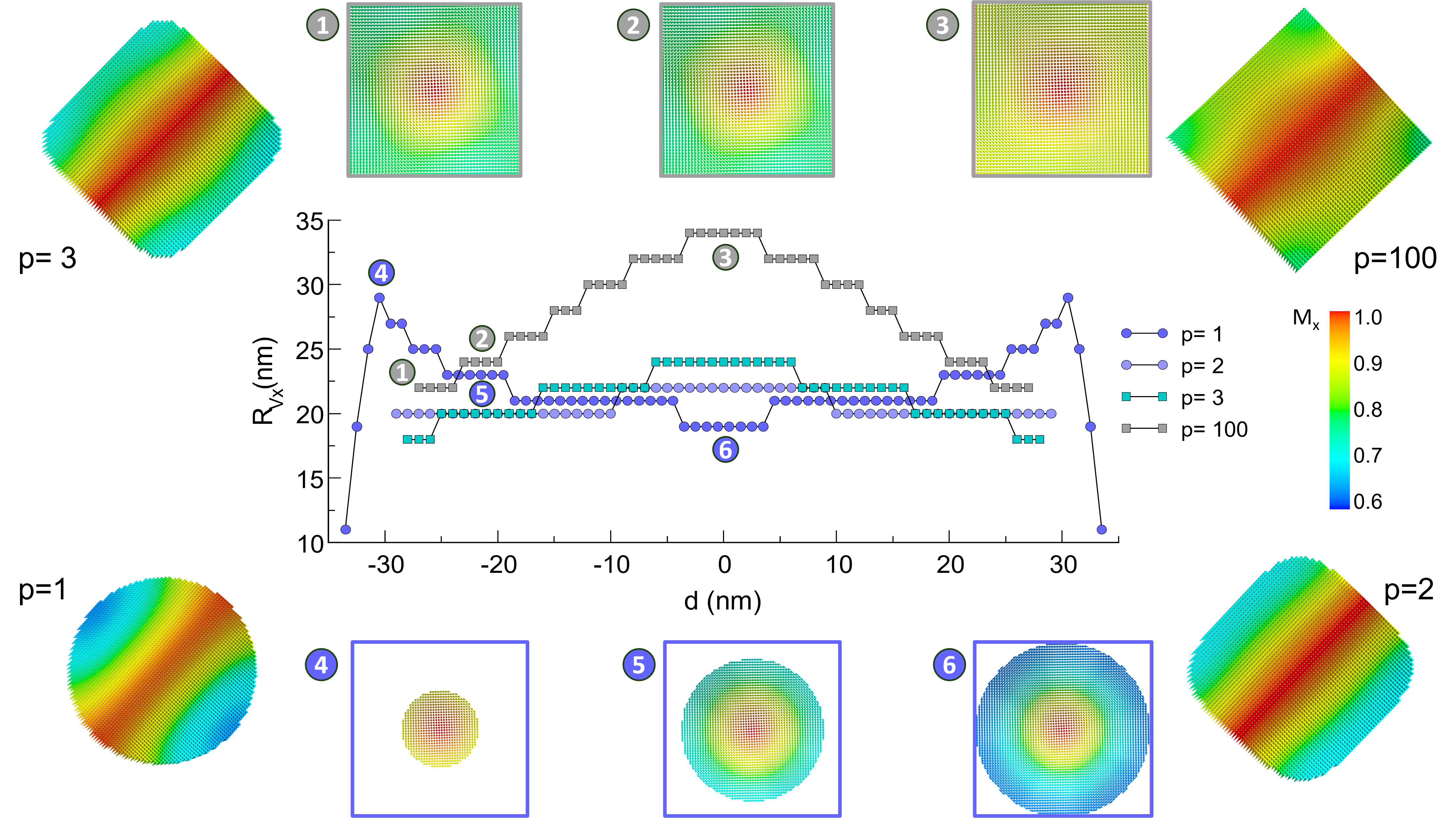}
\caption{\label{Fig_Config55} Vortex radius $R_{Vx}$ measured across different slices of NPs with cubic anisotropy and aspect ratio $r=1.0$. Results are shown for four NP characterized by shape exponents $p=1, 2, 3, 100$ and the same equivalent volume $V^{1/3}= 55$ nm.
Insets above and below the graph correspond to magnetization snapshots taken at the slice positions indicated by the numbered circles in the main plot. 
Snapshots located at the corners of the figure represent the central $yz$ plane section of each NP, showing the vortex core structure. Cone glyphs colours correspond to the normalized $M_x$ component according to the scale bars shown.  
}
\end{figure*}

As a representative case,  Fig.\,\ref{Fig_Config55} shows the magnetic configurations of NPs with different shape exponents $p= 1, 2, 3, 100$, all with approximate sizes of $55$ nm. This set corresponds to the symmetric case with an aspect ratio $r = $1.00, since the cases with \(r \ne\) 1.00 do not exhibit significantly different behaviors.
The snapshots at the corners of the figure, with cone glyphs representing the cell magnetizations on a $yz$ central plane passing through the middle of the vortex configurations, reveal striking differences in the magnetic ordering arising solely from variations in the NP shape.
For quasi-spherical NPs ($p= 1,2$) the magnetization curls around the $x$ axis (one of the hard anisotropy directions), with the $M_x$ component gradually decreasing from the vortex core toward the particle surface. 
This decrease is more pronounced for spherical NP, as indicated by dark blue regions near the surface in the lower left corner of Fig. \ref{Fig_Config55}, where $M_x\simeq 0.6M_s$.

In contrast, as the sphere deforms into a superball by increasing $p$, the drop in $M_x$ at the surface becomes less abrupt (see the shift to greener tones when increasing $p$ and the changes in the color bar scales). 
Simultaneously, the low $M_x$ regions first expand along the flat faces parallel to the vortex axis and subsequently migrate to the NP corners, which become more pronounced with increasing $p$.

Another notable feature concerns the inhomogeneous thickness of the vortex core along its axis, as highlighted by the reddish arrows in the figure. This non-uniformity appears consistently across all superball shaped NPs analyzed. 
The central panel of Fig.\,\ref{Fig_Config55}  shows the measured vortex radius as a function of the position $d$ along the vortex axis. The radius is defined as the distance where the $M_x$ component drops to $0.9M_s$. 
As seen in the plot, $R_{Vx}$ exhibits different behaviors depending on the shape of the NP. 
For spherical NP ($p=1$), the radius of the vortex core is smaller in the center than at the ends. For the superball shape with $p=2$, the vortex radius is relatively uniform along most of the NP, with slight maxima around the center. 
In contrast, as the particle becomes more cubic ($p=3$ and particularly $p=100$), the radius of the vortex becomes more strongly modulated, displaying a clear peak in the center of the NP and decreasing steeply toward the ends of the vortex line.

The top and bottom insets, corresponding to numbered cross-sections indicated in the plot, illustrate the local magnetization configuration in different slices. For spherical and nearly spherical NPs (bottom row, positions 4–6), the vortex core shows a gradual radial decrease of $M_x$, with circular symmetry and smooth variation in the color scale. The decrease in $M_x$ from the center to the surface is more pronounced for $p=1$, as indicated by the larger blue regions, signaling a lower axial magnetization. As $p$ increases, the $M_x$ component becomes more uniform, and the low $M_x$ regions shift toward the NP corners, in agreement with the increased localization, as seen in the vortex radius profiles.

The upper row of insets (positions 1–3) presents slices for highly faceted NP ($
p=100$), where the vortex structure is significantly more confined, with the core preferentially extending along the directions aligned with the faces and corners of the NP. This behavior reflects how the geometry-induced shape anisotropy in sharp-edged NPs modifies the spatial distribution of magnetization, redistributing low $M_x$ regions towards the NP corners and compressing the vortex core along the adjacent faces.
 
\begin{figure*}[!tp]
\centering
\includegraphics[width=0.9\textwidth]{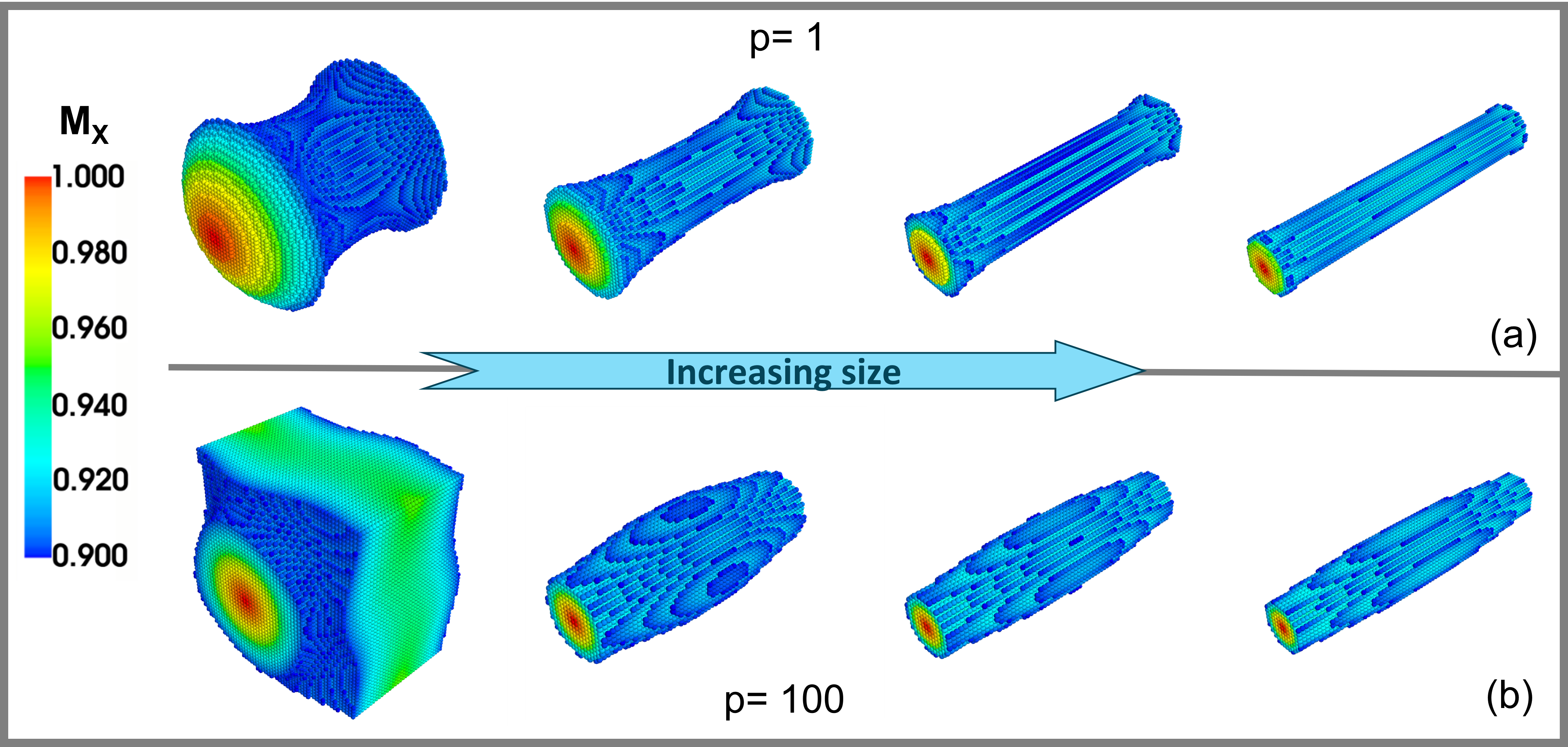}
\caption{\label{Fig_Vx_Sph}
Localized view of the vortex core in NPs with cubic anisotropy, highlighting regions where the axial magnetization component satisfies $M_x\ge 0.9M_s$. The configurations are visualized by displaying only the discretization cells (rendered as spheres) within this threshold, color coded according to their $M_x\ge 0.9M_s$ values. (a) Spherical NP with sizes $L = 52, 60, 70, 80$ nm (from left to right). (b) Cubic-shaped NP with increasing sizes $L = 61, 70, 83, 97$ nm (from left to right). 
}
\end{figure*}

To further elucidate how particle size and shape influence the internal vortex structure, Fig.\,\ref{Fig_Vx_Sph} presents a detailed view of the magnetization configuration in a narrow region surrounding the vortex core, where only cells with $M_x\ge 0.9M_s$ are shown. In all cases, the vortex remains centered within the nanoparticle, and any apparent displacement observed in some visualizations is solely due to the perspective used for the three-dimensional rendering.

In Fig.\,\ref{Fig_Vx_Sph}, for both cubic (panel a) and spherical (panel b) NPs, increasing particle size leads to a progressive elongation of the vortex core along the symmetry axis. 
However, the core morphology differs markedly as a result of the distinct shape-induced anisotropies. 
In cubic NPs, the vortex core assumes a faceted anisotropic structure that reflects the symmetry of the particle, with sharp edges and localized protrusions aligned with the cube faces. 
This confinement arises from the interplay between exchange interactions and the strong cubic shape anisotropy, which penalizes deviations of the magnetization away from the preferred directions near the flat surfaces and sharp corners. 

\begin{figure}[H]
\centering
\includegraphics[width=1.0\columnwidth]{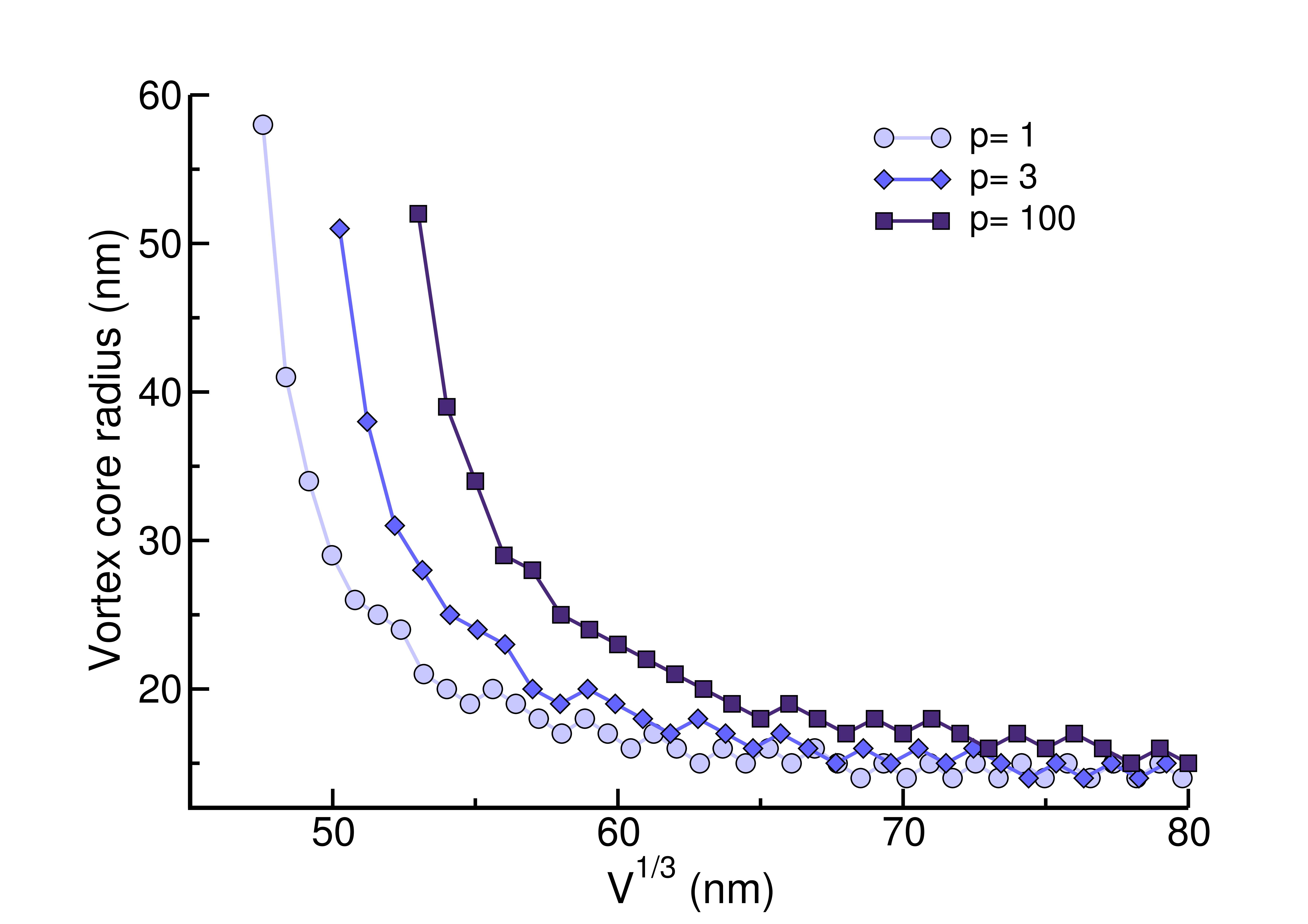}
\caption{\label{Fig_Vx_Rad}
Size dependence of the vortex core radius at the particle center, defined as the region where $M_x \geq 0.9 M_s$, for $p=1, 3, 100$.}
\end{figure}

As size increases, the core narrows laterally and stretches axially, suggesting an energy-favourable concentration of the vortex line to minimize surface demagnetization costs. 
In contrast, spherical NPs exhibit smoother, more isotropic vortex cores with radial symmetry and a more uniform decay of the $M_x$ component. 
The absence of sharp geometric constraints allows for a broader and more diffuse core structure that extends relatively evenly as the NP size increases. This comparison underscores how particle geometry not only affects the spatial extent of the vortex core but also modulates the balance between the different energy contributions, ultimately shaping the evolution of vortex morphology with size.
To conclude this analysis, and to complement the qualitative observations from Figs.~\ref{Fig_Config55} and \ref{Fig_Vx_Sph}, 
Fig.~\ref{Fig_Vx_Rad} provides a quantitative summary of the vortex core radius at the particle center as a function of size for three representative geometries ($p=1, 3, 100$). As seen, cubic nanoparticles exhibit a larger core radius than spherical ones for the same volume, in agreement with the trends already visible in Fig.~\ref{Fig_Vx_Sph}. In all cases, the vortex core radius decreases with increasing particle size, converging to values of 15 nm for sufficiently large volumes, independently of the particle geometry. 

\section{Conclusions}\label{sec_conclusions}
We have presented a comprehensive micromagnetic study of the influence of particle shape, magnetocrystalline anisotropy, and size on the magnetic configurations of magnetite NPs. 
Using the superball geometry, which allows for a continuous interpolation between spheres and cubes, we have systematically explored how subtle variations in NP morphology modulate equilibrium magnetization states and critical sizes for the quasi-uniform to vortex transition.
Particles with more faceted shapes (higher shape exponent values $p$) and higher axial ratios lead to increased critical volumes for the onset of vortex states.

Inclusion of cubic magnetocrystalline anisotropy, particularly relevant for magnetite, 
reduces this critical size across all shapes by favoring magnetic moment alignment along the [111] easy axes. This highlights the competition between shape-induced and intrinsic anisotropies in dictating the ground-state configuration of nanoscale magnets.

Our results underscore the non-trivial role of geometry-induced shape anisotropy in stabilizing uniform magnetic configurations. These configurations underpin the macrospin approximation commonly employed to model the magnetization dynamics of nanoparticle ensembles. By systematically analyzing how shape and size affect the onset of non-uniform states, our study provides a solid basis to delineate the range of validity of this approximation.

In-depth analysis of the magnetic configurations above the critical size has shown that the morphology of the vortex states is strongly affected by the NP geometry. We have shown that the vortex core is more compressed in the center than near the ends, whereas this trend reverses as the NP becomes more cubic. 
Furthermore, we show that even small deviations from the sphericity or ideal aspect ratio significantly influence the magnetic behavior of the NPs. Elongation or flattening of NPs introduces uniaxial shape anisotropy that competes with the intrinsic magnetocrystalline anisotropy, altering the stability and symmetry of magnetic states. 
These findings are particularly relevant in experimental contexts where particle monodispersity and idealized shapes are difficult to achieve, and highlight the need to account for realistic shape distributions to predict reliable results when modeling NP ensembles. As future work, it would be of interest to extend the particle shapes to include concaves ones such as recently reported in \cite{matsuo2025magnetization}. 
We plan to include applied magnetic fields to see how they affect the determination of the critical volumes, and also to consider thermal effects to bridge our findings with realistic experimental operating conditions.

\section*{Acknowledgements}
We acknowledge financial support by Spanish Ministerio de Ciencia, Innovaci\'on y Universidades through projects PID2019-109514RJ-100, PID2021-127397NB-100, and CNS2024-154574, "ERDF A way of making Europe", by the "European Union", and Catalan DURSI (2021SGR0032). Xunta de Galicia is acknowledged for projects ED431F 2022/005 and ED431B 2023/055. AEI is also acknowledged for  the \textit{Ram\'on y Cajal} grant RYC2020-029822-I that supports the work of D.S. We acknowledge the Centro de Supercomputacion de Galicia (CESGA) for computational resources.

\bibliographystyle{elsarticle-num}
\bibliography{Superspheres_2024}

\end{document}